\documentclass[conference]{IEEEtran}
\IEEEoverridecommandlockouts
\usepackage{cite}
\usepackage{amsmath,amssymb,amsfonts}
\usepackage{algorithmic}
\usepackage{graphicx}
\usepackage{textcomp}
\usepackage{xcolor}
\def\BibTeX{{\rm B\kern-.05em{\sc i\kern-.025em b}\kern-.08em
    T\kern-.1667em\lower.7ex\hbox{E}\kern-.125emX}}



\makeatletter
\newcommand*\titleheader[1]{\gdef\@titleheader{#1}}
\AtBeginDocument{%
	\let\st@red@title\@title%
	\def\@title{%
		\bgroup\normalfont\large\centering\@titleheader\par\egroup
		\vskip0.1em\st@red@title}
}
\makeatother

\title{ZSM-based Management and Orchestration of 3GPP Network Slicing: 
	An Architectural Framework and Deployment Options}
\titleheader{\textit{\textbf{This work has been submitted to the IEEE for 
			possible publication. \\Copyright may be transferred without 
			notice, after 
			which 
			this version may no longer be accessible.}}}

\begin{document}


\author{Muhammad~Mohtasim~Sajjad\textsuperscript{*}, 
	Dhammika~Jayalath\textsuperscript{*}, 
	Yu-Chu~Tian\textsuperscript{\dag}, 
	and~Carlos~J.~Bernardos\textsuperscript{\textsection}\\
	\textsuperscript{*}School of Electrical Engineering and Robotics,\\
	\textsuperscript{\dag}School of Computer Science,\\
	\textsuperscript{*}\textsuperscript{\dag}Queensland University of 
	Technology, Brisbane, Australia\\
	\textsuperscript{\textsection}Universidad Carlos III de Madrid, Spain  
	}%

\maketitle
\begin{abstract}
	Driven by closed-loop automation, the Zero-Touch Network and Services 
	Management (ZSM) framework offers invigorating features for the end-to-end 
	management and orchestration of a sliced network. Although ZSM is 
	considered a promising framework by 3GPP, there is a lack of concrete 
	ZSM-based solutions for the 3GPP Network Slicing management. 
	This article 
	presents an architectural framework of a ZSM-based management system for 
	3GPP Network Slicing. The proposed framework employs recursive ZSM 
	management domains to meet the specific management requirements of the 
	standard 3GPP Network Slicing framework. Two deployment options are 
	considered for 
	the presented framework. The first option features the integration of the 
	standard services of the 3GPP Management System within ZSM. The other one 
	considers ZSM as a complementary system for the standard 3GPP Management 
	System. From these deployment options, some key questions are identified on 
	ZSM’s interoperability with the existing 3GPP Network Slicing systems. 
	Finally, 
	for each option, the architectural and operational feasibility of the ZSM 
	interoperation with 3GPP Network Slicing systems is shown through an 
	example use case. 
\end{abstract}

\begin{IEEEkeywords}
ZSM, 5G/6G, 3GPP Network Slicing framework, Management and orchestration, 
Artificial Intelligence
\end{IEEEkeywords}

\section*{Introduction}
Management and orchestration of 3GPP Network Slicing brings complex challenges. 
Indeed, with the inherent flexibility of deployment offered by the 
Service-Based Architecture (SBA) \cite{23501}, the standard 3GPP Network 
Slicing framework offers necessary features to support highly diverse 
services and use cases. In addition to multi-service and multi-tenancy – the 
hallmark attributes of network slicing – the real-world deployment of network 
slicing would exhibit multiple and distinct administrative and technological 
domains \cite{5gpppWp}. Supporting such diversity over the often shared 
physical 
infrastructure, imposes stringent requirements for the management and 
orchestration of network slices. The traditional management systems, 
which include the standard 3GPP Management System 
\cite{28533}, are inadequate to meet such requirements. 

Zero-Touch Network and Service Management (ZSM) framework, proposed by the 
European Telecommunications Standards Institute (ETSI), offers the desired 
capabilities for effective end-to-end (E2E) management and orchestration of 
complex network 
slicing environments. Its primary aim is to achieve ideally 100 percent 
automation of management and orchestration tasks by employing different 
automation techniques 
including Artificial Intelligence (AI) \cite{Benzaid2020}. It has also been 
recognized among the key enablers for future 6G networks \cite{Jiang2021}. Due 
to 
the potential 
of ZSM, the 3GPP also supports delivering its standard management services 
through ZSM \cite{28533}. While a high-level framework for ZSM as a management 
framework 
for 3GPP Core Network (CN) and Radio Access Network (RAN) is given in 
\cite{28533}, 
there is a lack of concrete ZSM-based solutions which can meet the specific 
management requirements of the standard 3GPP Network Slicing 
framework.

This article presents an architectural framework of ZSM-based management system 
for 3GPP Network Slicing. The proposed solution is based on recursive ZSM 
management domains \cite{zsm002}. Apart from inheriting the ZSM benefits, the 
presented framework can promise the specific management needs of the standard 
3GPP Network Slicing framework. Specifically, the proposed architectural 
framework promises different degrees of \textit{management isolation} for 3GPP 
network 
slices and enables \textit{scalability} of the management system 
– both of which are recognized as the primary requirements of network 
slicing management systems \cite{Yousaf2019}. 

\subsection*{Terminologies}

Considering the commonly understood three-layered architecture for network 
slicing (comprising the \textit{management}, \textit{network slice instance}, 
and 
\textit{resource/infrastructure} layers), we have used the following 
terminologies 
throughout this article. The term \textit{3GPP Network Slicing systems} 
generically 
refers 
to systems at either of the three layers. These systems include the standard 
3GPP Management System at the \textit{management layer}, and the standard 3GPP 
network 
slice instances at the \textit{network slice instance layer}. At the 
\textit{resource/infrastructure layer}, the ETSI-specified Network Functions 
Virtualization (NFV) 
systems are considered since these are the widely used systems in most network 
slicing architectures.  The term \textit{standard 3GPP Network Slicing 
framework} refers 
to the standard framework of network slice instances as specified by the 3GPP. 
However, the management of the standard 3GPP Network Slicing framework, 
inevitably involves the management and orchestration operations at the 
resource/infrastructure 
layer, in addition to the management operations of the network slice instances. 
The term \textit{3GPP Network Slicing} is used as an abbreviated form of the 
standard 3GPP Network Slicing framework, and is used interchangeably with the 
latter. Finally, the E2E 3GPP Slice refers to a 3GPP network slice instance 
comprising (potentially multiple) RAN, Transport Network (TN), and CN 
domain(s). 

\subsection*{Organization}

The rest of this article is organized as follows. The next section presents a 
brief 
overview of the recent advances in ZSM-based management and orchestration of 
network slicing. The proposed ZSM-based architectural framework 
for 3GPP Network Slicing management is presented in the 
subsequent section. Two deployment options of the proposed framework with the 
standard 3GPP Management System are then introduced. The first option features 
the integration of the standard 3GPP Management services into ZSM. The other 
option considers deploying ZSM as a complementary system with the current 3GPP 
Management System. This section also discusses some open questions concerning 
the ZSM’s interoperability with the existing systems in the 3GPP 
Network Slicing. The following section then highlights the architectural and 
operational interoperability of the proposed framework with the aforementioned 
3GPP Network Slicing systems. An example, Virtualized Network 
Function (VNF) scaling process, is taken as a use case to describe the 
architectural and operational aspects in both deployment options. The final 
section concludes the article.

\section*{Related Work}\label{sec2}
Supporting multi-domain E2E management and orchestration
through AI-based closed-loop automation is the cornerstone
feature of ZSM.  
An AI-based cross-system management and 
orchestration solution in Reference
\cite{Bagaa2021} adopts 
\textit{Distributed Artificial Intelligence} techniques. Such techniques, which 
include transfer learning and federated learning etc., employ 
\textit{intelligent 
agents} in different management domains. Seamless 
collaboration 
between 
these agents enables the E2E network and service management across different 
domains. In Reference \cite{Xie2020}, collaboration among closed-loops in 
management 
domains is investigated to enable E2E service management in a
multi-provider, multi-vendor, and multi-tenant environment. 

Some solutions focus on enriching ZSM with additional features. For example, 
Reference \cite{Benzaid2020Security} proposes a security enhancement to the ZSM 
architecture, addressing its various identified security 
threats. 
The 
work 
in \cite{Rezazadeh2021} introduces the concept of \textit{Knowledge Plane (KP)} 
in ZSM. 
As a 
pervasive system, the KP provides high-level models of the network’s overall 
functioning. Based on KP, other elements in the network receive services and 
advice about their role and operation. To deal with trust and reliability 
issues for entities which belong to different administrative domains, 
block-chain technology is adopted within the ZSM framework in 
\cite{Theodorou2021}. 

\subsection*{Our Contribution}
Despite the current progress, the adoption of ZSM for the standard 3GPP Network 
Slicing 
framework has 
not 
been sufficiently explored. Apart from its inherent capabilities, ZSM for the 
standard 3GPP 
Network Slicing framework is also required to meet its specific management 
requirements. More specifically, the capability to promise management isolation 
among 
slices is critical since 
the standard 3GPP 
Network Slicing framework supports NFs sharing among different slices, in 
addition to the dedicated NFs. Furthermore, the scalability of ZSM systems is 
also 
essential for such 
deployments since 
continual provisioning and management of massive slices is expected in 
real-world scenarios \cite{23501}. 
The next section presents our proposed architectural framework for ZSM-based 
management of 3GPP Network Slicing that can ensure these capabilities.

\begin{figure*}
	\centering
	\includegraphics[scale=0.2]{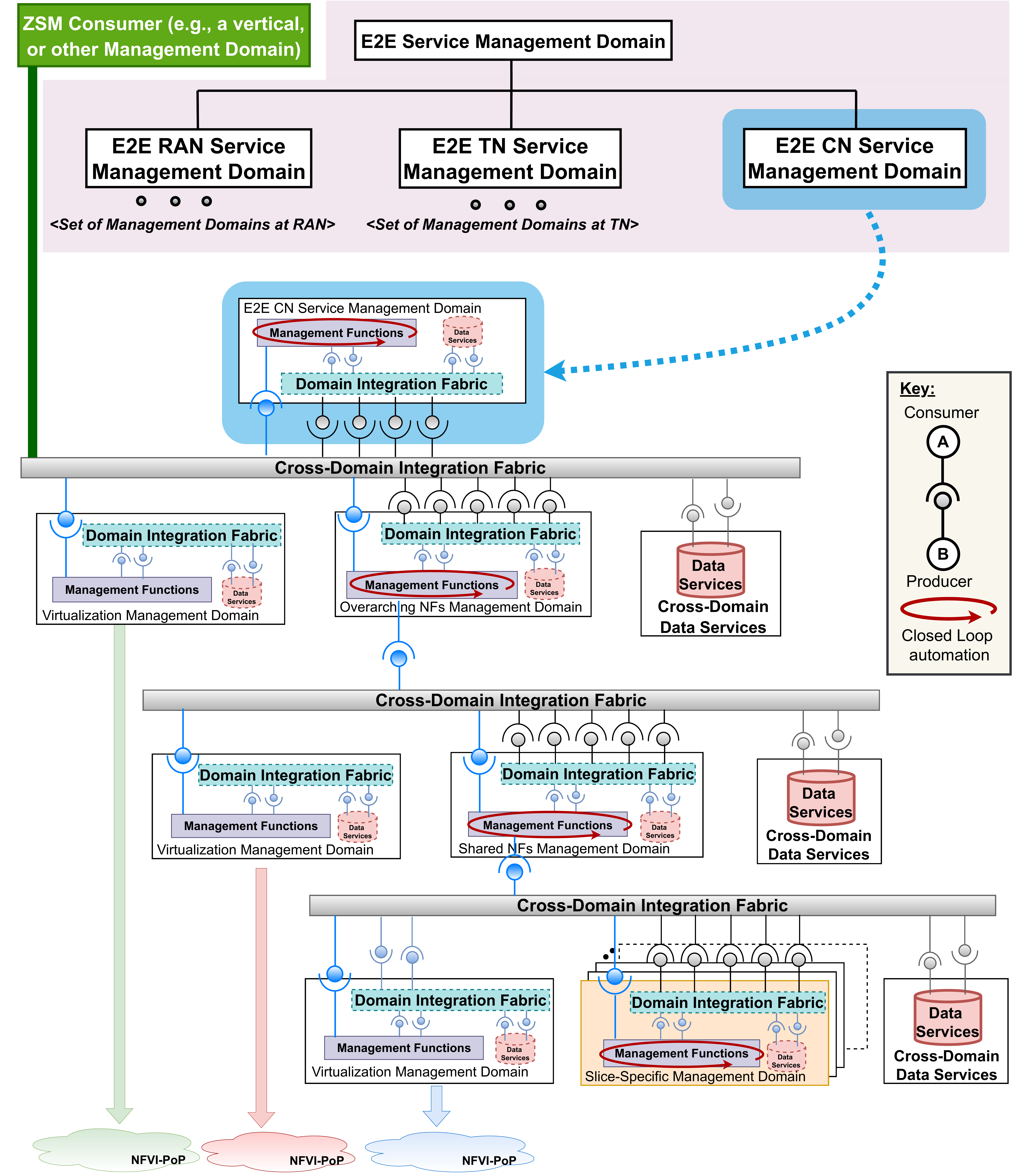}
	\caption{The proposed architectural framework.}
	\label{fig:ZSMSolutionWithE2E17}
\end{figure*}

\section*{Proposed Architectural Framework}
In this section, we present the proposed architectural framework of ZSM-based 
management system for 3GPP Network Slicing. The management isolation and system 
scalability demand a modular ZSM management system – a system exhibiting a high 
granularity of management domains, in addition to granularity in the management 
functions/services. As per the current ETSI ZSM specifications, such 
modularisation and granularity can be achieved by applying the ZSM 
architectural option of recursive management domains \cite{zsm002}.

In this option, a ZSM Management Domain, or a set thereof, is composed of other 
Management Domains \cite{zsm002}, over which an E2E service can be deployed. In 
essence, recursion enables the creation of multiple (possibly infinite) 
management 
domains, each with varying degrees of self-containment and self-dependence. 
Depending on the management and orchestration needs of the underlying domain, 
the management 
functionalities of different management domains can be similar or completely 
different from each other. As 
noted in \cite{zsm002}, the option of recursive deployments is also applicable 
for the E2E Service Management Domain. Such deployments become useful for 
managing slices that may cover very wide coverage areas, and be potentially 
composed of 
several technological and administrative domains. Our proposed framework 
featuring the recursive ZSM management domains is shown in Figure 
\ref{fig:ZSMSolutionWithE2E17}.

In the proposed framework, the E2E service management domain, responsible for 
the E2E service management and orchestration, is composed of E2E RAN, E2E CN, 
and E2E TN service management domains. Such architecture 
allows the deployment and management of service instances independently over 
RAN, CN and TN domains, which have distinct technological and topological 
characteristics \cite{Chahbar2021}. To focus on the interoperability of ZSM 
with 3GPP Network 
Slicing systems and also for simplicity, we only consider the CN 
domain for 
3GPP Network Slicing henceforth. The control plane of the 3GPP CN domain 
constitutes SBA, and hence we use the terms SBA and CN control plane 
interchangeably.

The CN control plane NFs which can be 
optionally shared among different slices include Access and Mobility Management 
Function (AMF) and Network Repository Function (NRF). In comparison, the NFs 
dedicated to a slice (or slice-specific NFs) in any typical deployment include 
Session 
Management Function (SMF), Policy Control Function (PCF)
and Network Data Analytics Function (NWDAF). In addition, a set of overarching 
NFs at CN, which are common to 
all slices, also exist which includes the Network Slice Selection Function 
(NSSF). The instances of other NFs, such as Unified Data Management (UDM), Unified 
Data 
Repository (UDR), and Unstructured Data Storage Function (UDSF), can be 
slice-specific, common, and overarching, depending on the deployment needs. 
Correspondingly, 
in the proposed framework, three levels of CN management domains are 
defined for \textit{slice-specific}, \textit{shared}, and \textit{overarching} 
NFs of the slice, respectively. 

The modularization of ZSM eases its programmability. It enables 
the provisioning of \textit{tailored}, and \textit{slice-specific} management 
domains when 
a request 
for slice/service 
creation is received from a ZSM Consumer (e.g., a vertical). A 
\textit{tailored} 
management domain provides the slice owners (e.g., the verticals) with 
distinct options for management. It allows them to enforce their desired 
policies as well as 
control the management services exposure and data sharing towards the ZSM 
\textit{Cross-domain data services}. Hence, despite sharing some NFs with other 
slices, management isolation to slice owners can be promised. In addition, a 
modularized ZSM system also facilitates the independent addition of new 
management 
domains at any level as new slices are added, thus making the ZSM system 
scalable. In essence, the modularized ZSM system, with management 
isolation and 
scalability attributes, becomes an enabler for autonomous management and 
orchestration operations in individual domains constituting the E2E 3GPP 
slice. 
Similar architecture 
designs based on recursive ZSM management domains may be considered for 
multi-domain RAN and TN architectures as well. They are, however, omitted here. 



The NFV-based virtualization environments (e.g., NFVI-PoPs) have demonstrated 
suitability for deploying VNFs  
constituting a slice. For such deployments, apart from regular ZSM 
management domains, Virtualization management domains are also necessary. A 
Virtualization management domain for NFV employs the standard NFV Management 
and 
Orchestration (NFV-MANO) components as ZSM 
services. It is responsible for \textit{resource-level} management of VNFs. 
That is, 
the compute, networking, and storage resources of, for instance, the 
slice-specific NFs 
(in Figure \ref{fig:ZSMSolutionWithE2E17}) will be managed by the 
Virtualization management domain. 
However, any \textit{application-level} interaction of NFs such as NWDAF and 
PCF will still 
occur through the regular Management Domain services.
Note that
in contrast to multiple Virtualization management domains for different 
NFVI-PoPs shown in Figure \ref{fig:ZSMSolutionWithE2E17}, a single 
Virtualization management domain 
would suffice for deployments where all NFs of a slice can be hosted at a 
single 
location (e.g., alongside the Overarching NFs management domain).

\begin{figure*}
	\centering
	\includegraphics[scale=0.6]{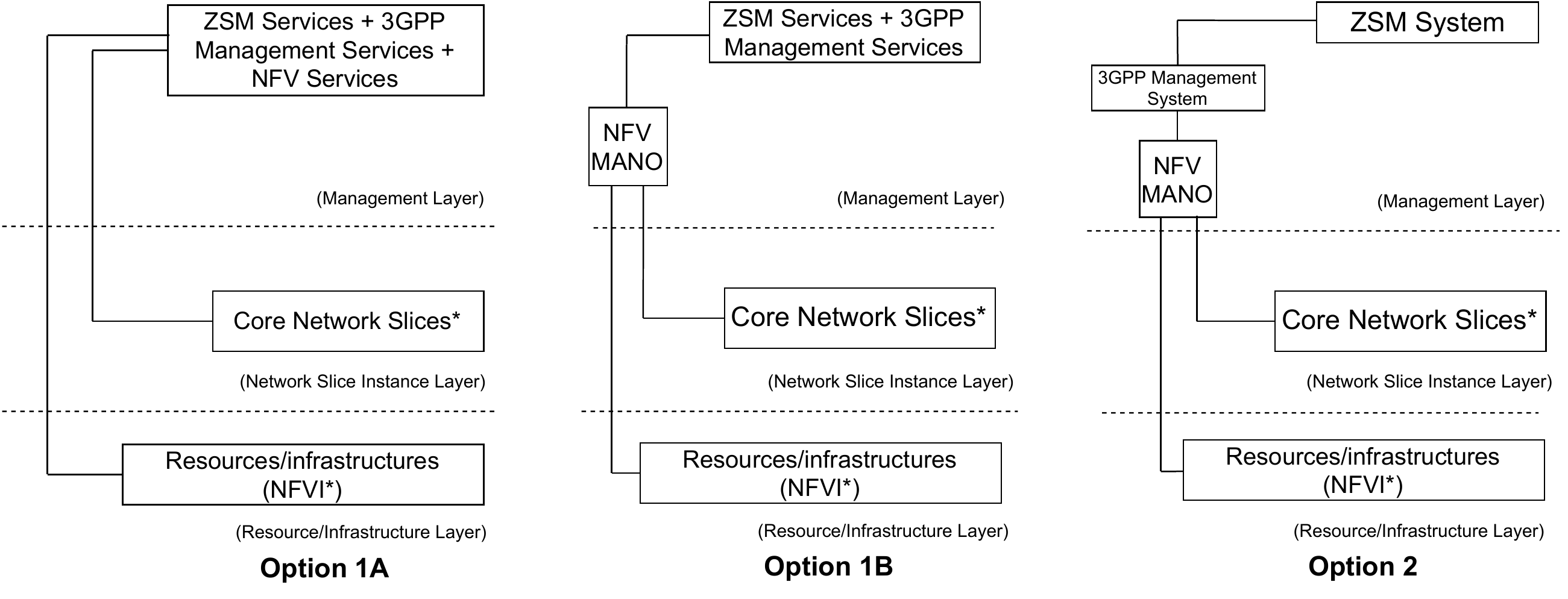}
	\caption{Deployment Options: Integrated (Option 1A and 1B), and 
		Complementary (Option 2) \textit{(*Assuming NFV-based control plane NFs 
			deployment only).}}
	\label{fig:BlockDIAGRAMS01}
\end{figure*}

\section*{Deployment Options}
The ongoing standardization efforts for ETSI ZSM are still at an early stage. 
Their 
specifications are preliminary, and are based on a higher level of abstraction. 
In this regard, 
envisioning the ZSM deployment scenarios in the real-world slicing environments 
is essential, as it can foster a practical way forward for identifying and 
addressing other challenges and providing their practical and viable solutions. 
%
%
%
%
%
Hence, having described its ability to promise management isolation and 
scalability, we now focus on the deployment options for the proposed framework. 

The first deployment 
option features the integrated ZSM and 3GPP Management systems. This option is 
consistent 
with 
3GPP’s conceptualization of ZSM system for 3GPP Network Slicing management 
\cite{28533}. 
Generally, such deployments will also incorporate an integrated NFV-MANO 
within ZSM \cite{zsm002} (Option 1A in Figure \ref{fig:BlockDIAGRAMS01}). 
However, standalone NFV-MANO deployments with the 
integrated 
ZSM and 3GPP Management System are also possible \cite{zsm004} (Option 1B in 
Figure 
\ref{fig:BlockDIAGRAMS01}). In Option 1A, 
the 3GPP Management services, and 
NFV-MANO functionalities become part of the ZSM services set. Hence, services of
either system can be mutually invoked by authorized services of any other 
system. However, for Option 1B, the inter-operation of standalone NFV-MANO with 
the integrated ZSM and 3GPP Management System brings some interworking 
challenges, which will be
highlighted later. 

The higher modularization of ZSM management domains are expected to exhibit 
multiple 
closed-loops for automation at different levels of hierarchy. This would result 
in high computational complexity. Accordingly, its feasibility to 
provide a complete set of management services for large-scale and massive 
number of 
slices becomes challenging. In this regard, the operator may wish to utilize
ZSM services 
selectively for chosen slices only. That is, the ZSM system may only act as a 
complementary system to the existing 3GPP Management System. In such 
deployments, ZSM, 3GPP Management System, and NFV-MANO will act as standalone 
systems interacting with each other through suitable interfaces. This is 
represented as Option 2 in Figure \ref{fig:BlockDIAGRAMS01}. 
\subsection*{Open Questions}
For each of the aforementioned deployment options, certain questions arise on 
ZSM’s 
interworking with the existing 3GPP Network Slicing systems. Firstly, some ZSM 
services  
represent an overlap with the standard 
3GPP control plane NFs at the underlying slice instances. Secondly, interfacing 
challenges of ZSM arise with existing systems including the standalone 3GPP 
Management System, the standalone NFV-MANO and slice instances at the 3GPP CN 
control plane. We 
discuss these in the following.

%
%
%
%

\subsection*{Overlap of ZSM Services with SBA NFs}
\begin{enumerate}
	\item \textit{Data Analytics:} 
	The NWDAF provides data analytics functionality at the SBA. 
	This represents an overlap with the ZSM Data Analytics 
	service. However, by distributing the data 
	analytics functionalities among the control plane and the management plane, 
	such 
	overlap can be averted. As 
	discussed in \cite{Bega2020}, such approaches can enable the deployment of 
	AI-based 
	algorithms for short-term forecasting at the control plane, while 
	medium-term, and long-term forecasting at ZSM. 
	
	\item \textit{Data Storage: }
	The 3GPP specifications have defined control plane NFs such as UDSF, UDM 
	and UDR, 
	which are responsible for data collection and storage \cite{23501}. This 
	also 
	represents an overlap 
	with ZSM Data Storage services. A ZSM system managing the slice 
	requires constant, and up-to-date data retrieval from these NFs. Yet again, 
	the data required for localized functionalities (e.g., data analytics at 
	control 
	plane) may only be stored at control plane 
	resources, while the rest be transferred to the management plane 
	entities. 
	\item \textit{Policy Management: }
	The ZSM Policy Management service and the 3GPP PCF are both defined for 
	policy 
	management related functionalities. With the interaction of control and 
	management planes, it is possible to deploy static policies at the control 
	plane 
	through 
	PCF. On the other hand, dynamic policies can be generated, for instance, 
	through ZSM Domain Intelligence services, and then intermittently 
	transferred to PCF for 
	enforcement. 
\end{enumerate}

\subsection*{Interfacing ZSM with Existing Systems}
\begin{enumerate}
	\item \textit{CN Control Plane NFs: }
	For functionalities like data collection and other interactions, suitable 
	interfacing between ZSM and SBA NFs is required. Since 
	both 
	ZSM and SBA are service-based, enhancing an SBA NF with Management Function 
	interfaces as 
	specified in \cite{28533} will enable such interaction. 
	
	\item \textit{Interaction with 3GPP Management System:} For Option 
	2 (Figure \ref{fig:BlockDIAGRAMS01}), where the 3GPP Management System 
	exists as an external, 
	standalone 
	system, mutual services exposure from both systems is required for their 
	interworking. The ZSM 
	specifications
	have defined the \textit{ZSM Management Services Adapter} for this purpose 
	\cite{zsm002}. 
	Similar 
	functionality from 3GPP Management System side is performed by the 
	\textit{Exposure Governance Management Function (EGMF)}. Since both ZSM and 
	3GPP 
	Management System are 
	service-based, such interaction can be realized with 
	fewer challenges. 
	\item \textit{Interaction between NFV-MANO and ZSM:} The standard NFV-MANO 
	interfaces are not service-based. This is a major challenge for ZSM's 
	interaction 
	with the standalone NFV-MANO, as indicated in \cite{zsm004}. The 
	\textit{ZSM 
	Management 
	Service Adapters} to enable such interaction are required to 
		translate the 
		NFV-specific communication into ZSM and vice versa, and as such are 
		expected to be 
		more 
	sophisticated. 
	\item \textit{Other systems:} 
	At the resource/infrastructure layer, in addition to the VNFs,  elements 
	such as 
	physical 
	NFs and containerized NFs are also possible, which collectively constitute 
	a slice. These elements will also 
	require ZSM specific interfaces to interact with the ZSM 
	System. 
	
\end{enumerate}

\section*{VNF Scaling Use Case}
This section aims to delve deep into the technical aspects of the 
aforementioned deployment options, highlighting key services of the involved 
systems and their interworking. The process of VNF scaling is taken as a 
use case to describe both options. We choose this process since it can 
be triggered to meet both service management  and the 
slice management requirements. 
%
%
%
%
%
%

\subsection*{Integrated Option}
\label{sec:IntegratedOption}
\begin{figure}
	\centering
	\includegraphics[scale=0.135]{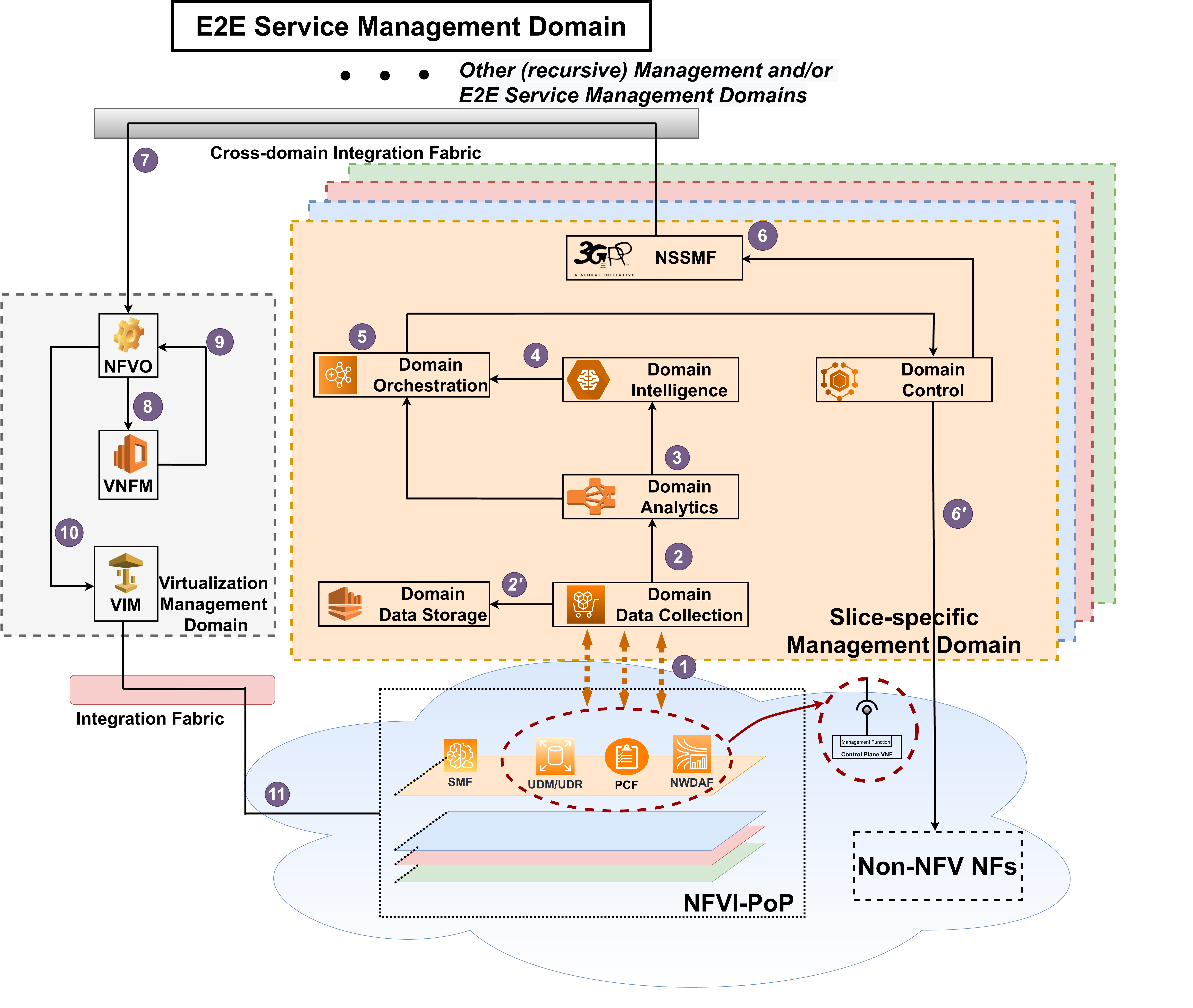}
	\caption{Integrated option (Option 1A) showing deployment of 3GPP 
		Management and NFV-MANO services within ZSM.}
	\label{fig:IntegratedSolution}
\end{figure}

We assume that the CN control plane NFs including PCF, UDM/UDR, UDSF, and NWDAF are 
enhanced with 3GPP Management Functions interfaces \cite{28533} as shown in 
Figure \ref{fig:IntegratedSolution}. The ZSM \textit{Domain Data Collection} 
services can receive 
the continuous data streams from these NFs \textbf{(1)}. The \textit{Domain 
Data Collection} may choose to store the received data in the \textit{Domain 
Data 
Storage} \textbf{(2$'$)}, or send it to the \textit{Domain Analytics} 
\textbf{(2)}. 
Based on the received data, 
and some additional information (e.g., of control plane topology and allocated 
resources to each NF), the \textit{Domain Analytics} service can generate 
insights and 
predictions \cite{zsm002} within the purview of the current management domain. 
For 
instance, the \textit{Domain Data Analytics} and within these the 
\textit{Anomaly detection 
service} can detect anomalies in response time of a particular CN control plane 
NF at the slice-specific instance of the current slice \textbf{(3)}. The 
\textit{Domain Data Analytics} then invokes the \textit{Domain Intelligence} 
services, which are 
responsible
for deciding if any actions are required. 

By employing AI techniques and running the 
closed loop automation, the \textit{Domain Intelligence} services infer the 
need for VNF scaling to remedy the detected 
anomaly \textbf{(4)}. 
Once the decision is made, the planning for action(s) is done. Such actions 
include 
Orchestration/Control actions, which are to be executed by the ZSM services 
from \textit{Domain Orchestration} and \textit{Domain Control} services.
The Orchestration services, based on the available \textit{domain service 
model}, which contains the 
complete description of infrastructure resources, can determine the feasibility 
and extent of VNF scaling \textbf{(5)}. 

The modification of resources at the resource/infrastructure layer is executed 
by the 
\textit{Domain Control} services. 
Subsequent steps would normally involve the interaction of \textit{Domain 
Control} services with the underlying infrastructure \textbf{(6$'$)}. However, 
in the 
integrated option, Option 1A, the \textit{Domain Control} services are required 
to 
invoke the NSSMF services. 
For 
VNF scaling, the \textit{Resource Lifecycle Management} service from 
\textit{Domain 
Control} 
services invokes the \textit{Network 
Slice Subnet Provisioning Management} service of the integrated NSSMF. This 
NSSMF service is defined in the 3GPP Technical Specification, \textit{TS 
28.531} 
\textbf{(6)}. 
If 
	the 
NSSMF 
determines that the VNF scaling is possible, the NSSMF instance invokes the
\textit{NFV Orchestrator (NFVO)} services from the Virtualization management 
domain 
\textbf{(7)}. 

The NFVO will execute the VNF scaling (as per \textit{ETSI Group Specification, 
ETSI GS 
NFV-MAN001}) as follows. After 
validating the received VNF Scaling Request, and performing the feasibility 
check, the NFVO sends the Scaling Request to the relevant \textit{VNF Manager 
(VNFM)}. 
The 
VNFM may perform certain preparation tasks (e.g., evaluating the request as 
per VNF lifecycle constraints) \textbf{(8)}. Next, the VNFM invokes the 
Scale 
Resource operation of the NFVO \textbf{(9)}. The NFVO sends a request to 
\textit{Virtualized Infrastructure Manager (VIM)} to 
change 
resources (compute, storage, networking) required to scale the VNF. The VIM 
eventually modifies 
the resources to effectuate the resource scaling of the VNF \textbf{(10)}. 

\subsection*{Complementary Option}
In the complementary option, the authorized 3GPP Management Services, for their 
specific operational needs, can invoke the ZSM services. 
For instance, the 
NSMF, through \textit{ZSM Management Services Adapter} may invoke management 
services at 
the E2E Service management domain. The E2E Service management domain may in 
turn invoke the respective management domain's services to start collecting 
data from the underlying slices and performing the analytics. Note that the 
actual slice 
provisioning and configuration in such 
deployments will still be carried out by the 3GPP 
Management System. That is, the ZSM system will not feature any 
Virtualization management domains. The ZSM overhead can also be reduced through 
the
complementary option. For instance, the NSMF might subscribe to the 
\textit{Domain Analytics} service to receive 
notifications for fault or anomaly detection etc. The NSMF 
may then instruct the NSSMF to perform the \textit{Fault Management} operations 
to 
remedy the faults, thus relieving the ZSM system further from running the
closed-loop automation. 

\begin{figure}
	\centering
	\includegraphics[scale=0.125]{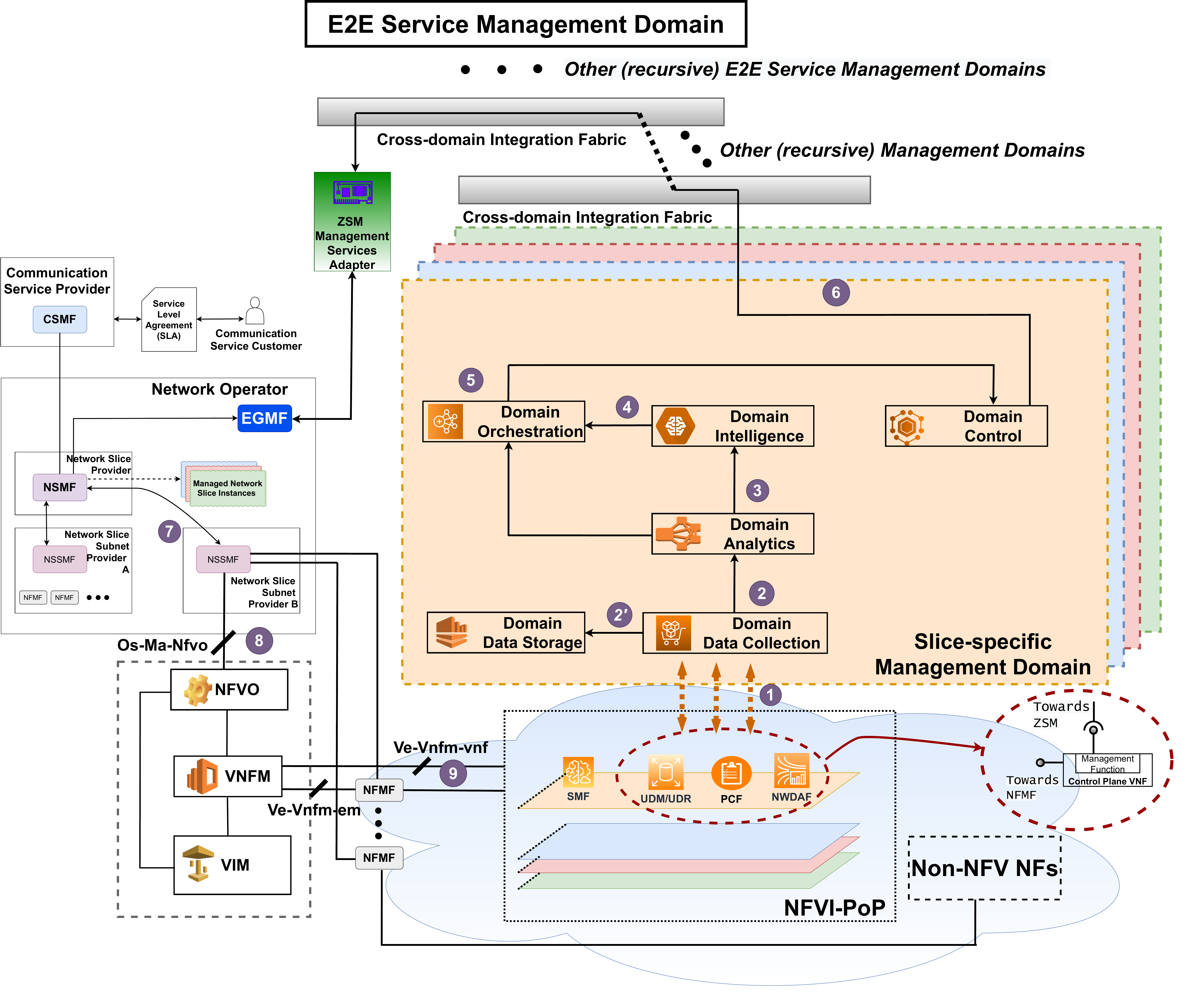}
	\caption{Complementary option (Option 2) showing deployment of ZSM for 3GPP 
		Management System.}\label{fig:ComplementarySolution}
\end{figure}

For VNF scaling operation, following the steps \textbf{(1)} to \textbf{(5)} of 
the Integrated Option (in previous sub-section), the \textit{Domain Control} 
services are required to invoke the external 3GPP Management 
System services (Figure \ref{fig:ComplementarySolution}). For VNF scaling, the 
NSMF's \textit{Network Slice Provisioning 
Management Service} exposed to the ZSM system can be invoked by the 
\textit{Domain 
Control} services \textbf{(6)}. The NSMF, in turn, invokes the \textit{Network 
Slice 
Subnet 
Provisioning Management Services} of the respective NSSMF \textbf{(7)}. NSSMF 
then utilizes the \textit{Os-Ma-nfvo} interface of NFV-MANO \cite{28533} to 
send the 
\textit{VNF Scaling} request to the NFVO \textbf{(8)}. The \textit{VNF 
Scaling} then takes place through VNFM and the VIM as described in the 
Integrated Option (in the previous sub-section) \textbf{(9)}. 

\section*{Conclusions}
ZSM is now considered an indispensable part of future management systems for 
network slicing. This article has investigated the applicability of ZSM in an 
operational 3GPP Network Slicing system. An architectural framework of ZSM 
based on 
recursive ZSM management domains is proposed. The deployment aspects of the 
proposed ZSM framework
with 3GPP Network Slicing have also been discussed. Two possible options for 
deployment are discussed, 
one featuring the integrated 3GPP Management System within ZSM, and the other 
considering ZSM as a complementary system to the 3GPP Management System. For 
each 
option, we have shown the architectural and operational feasibility of ZSM  
interoperation with the existing 3GPP Network Slicing systems through an 
example use case.

\bibliographystyle{elsarticle-num}

\begin{IEEEbiographynophoto}{Muhammad Mohtasim Sajjad} [S'17] 
(m.sajjad@ieee.org)
	is a PhD candidate at the Queensland University of Technology (QUT), 
	Brisbane, Australia. His research interests include network slicing, 
	network virtualization and mobile communications.  
\end{IEEEbiographynophoto}

\begin{IEEEbiographynophoto}{Dhammika Jayalath} [M'99, SM'11] 
(dhammika.jayalath@qut.edu.au)
	is a Senior Lecturer at the Queensland University of Technology (QUT), 
	Brisbane, Australia. His current research interests includes 5G systems, 
	cooperative communications, communications theory. 
\end{IEEEbiographynophoto}

\begin{IEEEbiographynophoto}{Yu-Chu Tian} [M'00, SM'19] (y.tian@qut.edu.au)
	is currently a Full Professor at QUT, Brisbane Australia. His current 
	research 
	interests include big data computing, cloud computing, computer networks, optimization and machine learning, networked control systems, and cyber-physical system security. 
\end{IEEEbiographynophoto}

\begin{IEEEbiographynophoto}{Carlos J. Bernardos} (cjbc@it.uc3m.es)
	is affiliated with the University Carlos  III  of  Madrid  (UC3M), Spain. 
	His current research focuses on cloud technologies, beyond 5G networks. 
\end{IEEEbiographynophoto}

\end{document}